\documentclass[fleqn,twoside]{article}
\usepackage{espcrc2}
\usepackage{cite}

\readRCS
$Id: espcrc2.tex,v 1.2 2004/02/24 11:22:11 spepping Exp $
\usepackage{epsfig}

\def\NNLO{{\mbox{NNLO (approx)}}}
\def\sigmaNNLO{\ensuremath{{\sigma_{\scriptstyle\NNLO}}}}
\def\NLO{{\mbox{NLO}}}

\def\mt{{m_t}}

\def\muf{{\mu^{}_f}}
\def\mufs{{\mu^{\,2}_f}}
\def\mur{{\mu^{}_r}}
\def\murs{{\mu^{\,2}_r}}

\newcommand{\central}{${\sigma(\mu=\mt)}$}
\newcommand{\pdfup}{${\sigma(\mu=\mt)+\Delta\sigma_{PDF}}$}
\newcommand{\pdfdown}{${\sigma(\mu=\mt)-\Delta\sigma_{PDF}}$}
\newcommand{\highscale}{${\sigma(\mu=2\mt)}$}
\newcommand{\lowscale}{${\sigma(\mu=\mt/2)}$}

\newcommand{\AmS}{{\protect\the\textfont2
  A\kern-.1667em\lower.5ex\hbox{M}\kern-.125emS}}

\title{
\vspace*{-26mm} \rightline{ {\normalsize{DESY 08--097}}}
\vspace*{-2mm} \rightline{ {\normalsize{TTP 08-29}}}
\vspace*{-2mm} \rightline{ {\normalsize{SFB/CPP-08-49}}}
\vspace*{-2mm} \rightline{ {\normalsize{July 2008}}}
\vspace*{+6mm}
Heavy-quark pair production at two loops in QCD%
{\thanks{Presented by S.M. at {\it{Loops and Legs in Quantum Field Theory}}, 
20--25 April 2008, Sondershausen (Germany).}}%
}

\author{S. Moch\address{DESY, Platanenallee 6, D--15738, Zeuthen, Germany}
and
P. Uwer\address{Institut f\"ur Theoretische Teilchenphysik, Universit\"at Karlsruhe, D--76128 Karlsruhe, Germany}
}

\begin{document}

\begin{abstract}
We present updated predictions for the total cross section of top-quark pair production at Tevatron and LHC.
For the LHC we also provide results at $\sqrt{s}$ = 10~TeV, in view of the anticipated run in 2008 
and quote numbers for the production of new heavy-quark pairs with mass in the range 0.5 -- 2~TeV.
Our two-loop results incorporate all logarithmically enhanced terms near threshold
including Coulomb corrections as well as the exact dependence on the
renormalization and factorization scale through next-to-next-to-leading order in QCD.
\end{abstract}

\maketitle

\section{Introduction}
\label{sec:intro}
Research on top-quark physics at hadron colliders has received great interest
in the past years in view of the steadily improving measurements at Tevatron and
the upcoming LHC (see Ref.~\cite{Bernreuther:2008ju} for a recent review).
In this respect, the total cross section for top-quark pair production is a quantity of great importance 
for experimental analyses and even allows for measurements of the top-quark mass.

Moreover, on the theory side, the total cross section has been subject to
numerous studies the motivation being improved predictions beyond the
long-known next-to-leading order (NLO) corrections in QCD~\cite{Nason:1988xz,Beenakker:1989bq,Bernreuther:2004jv}.
Recent work in this direction has aimed at completing the next-to-next-to-leading order (NNLO) QCD
predictions~\cite{Dittmaier:2007wz,Czakon:2007ej,Czakon:2007wk,Korner:2008bn,Czakon:2008zk},
at resumming large Sudakov logarithms to next-to-next-to-leading logarithmic accuracy~\cite{Moch:2008qy}
and, at estimating bound state effects~\cite{Hagiwara:2008df}.
Also our knowledge on the parton distribution functions (PDFs) and the precision of the
top-quark mass determination has continuously improved over the last years.

In order to study the impact of the various improvements on Tevatron and
LHC predictions we build on the recent results of Ref.~\cite{Moch:2008qy}.
These approximate NNLO results for the total cross section are based on the complete logarithmic dependence on the heavy quark
velocity $\beta = \sqrt{1-4m^2/s}$ near threshold $s \simeq 4m^2$. Moreover,
they include the complete two-loop Coulomb corrections 
as well as the exact dependence on the renormalization and factorization scale at NNLO~\cite{Kidonakis:2001nj}.

Recently, similar studies have appeared in Refs.~\cite{Kidonakis:2008mu,Cacciari:2008zb,Nadolsky:2008zw}. 
While Ref.~\cite{Kidonakis:2008mu} largely follows our approach~\cite{Moch:2008qy}
to describe the total top-quark pair cross section at NNLO, 
Ref.~\cite{Cacciari:2008zb} has limited itself to updating older predictions 
based on threshold resummation to next-to-leading logarithmic accuracy only.
Thus, Ref.~\cite{Cacciari:2008zb} necessarily arrives at larger theoretical uncertainties.
The interesting study of Ref.~\cite{Nadolsky:2008zw} on the other hand applied
consistently predictions to NLO accuracy in QCD. 
In doing so, it has investigated correlations of rates for top-quark pair production with 
many other cross sections at LHC to quantify a potential sensitivity to the gluon luminosity.

\section{Total cross section}
\label{sec:total}
The total hadronic cross section for top-quark pair production 
depends on the hadronic center-of-mass energy squared $s$ and
the top-quark mass $\mt$. It is given by 
\begin{eqnarray}
  \label{eq:totalcrs}
  \sigma(s, \mt^2) &=&
  \sum\limits_{i,j = q,{\bar{q}},g} 
  f_{i/p}\left(\mufs\right) \otimes
  f_{j/p}\left(\mufs\right) 
  \nonumber\\ & &
  \otimes \,\,
  \hat{\sigma}(\mt^2,\mufs,\murs)\, ,
\end{eqnarray}
where $f_{i/p}$ are the PDFs of the proton.
The partonic cross section is given by $\hat{\sigma}$ and 
$\otimes$ denotes the standard convolution (see e.g. Ref.~\cite{Moch:2008qy}).

The generally adopted procedure to estimate the theoretical uncertainty 
for $\sigma$ in Eq.~(\ref{eq:totalcrs}) exploits the residual dependence 
on the renormalization and factorization scale, $\mur$ and $\muf$, 
which are identified throughout this article (i.e. $\mur = \muf = \mu$).
The NLO QCD corrections for the parton cross section $\hat{\sigma}$ and the PDFs $f_{i/p}$
provide the first instance where a meaningful error can be determined in this way.
We define the range as 
\begin{eqnarray}
  \label{eq:range}
  {\lefteqn{
      \sigma(\mu=2\mt)-\Delta\sigma_{PDF}(\mu=2\mt) 
      \, \le \,
  \sigma(\mu)
    }}
  \nonumber \\ && 
  \, \le \,
  \sigma(\mu=\mt/2)+\Delta\sigma_{PDF}(\mu=\mt/2)
  \, ,
\end{eqnarray}
where $\Delta\sigma_{PDF}$ is computed from the variation of the cross section
with respect to the parameters of the global fit (see e.g. Refs.~\cite{Nadolsky:2008zw,Martin:2007bv,Tung:2006tb}).

In this contribution we employ the approximate NNLO result~\cite{Moch:2008qy} 
to predict cross sections~(\ref{eq:totalcrs}) and the associated uncertainty ranges~(\ref{eq:range}) 
at Tevatron and LHC. 
Let us therefore briefly comment on the anticipated accuracy.
Our cross section \sigmaNNLO\ takes along all logarithmically enhanced terms $\ln^k\beta$, $k=1,\dots,4$ 
as well as the complete Coulomb corrections ($\sim 1/\beta, 1/\beta^2$) 
at two loops for the dominant parton channels $q{\bar q}$ and $gg$ 
and adds them on top of the exact NLO predictions.
In this way, our predictions rely on exact expressions in the region of phase space $s \simeq 4m^2$, 
where perturbative corrections receive the largest weight from the convolution with
the parton luminosities, cf. Eq.~(\ref{eq:totalcrs}).
The effect of new parton channels opening at NNLO 
($qq$ and ${\bar q}{\bar q}$) is expected to be small, cf. the $qg$ and ${\bar q}g$ channels at NLO. 

The region of large energies $s \gg 4m^2$ on the other hand is inaccessible
within our approach~\cite{Moch:2008qy}.
However, it is expected to give only small contributions in a full NNLO
calculation in line with the observed small corrections for top-quark pair production
together with an additional jet at NLO~\cite{Dittmaier:2007wz} 
which are part of the full NNLO correction for top-quark pair production.

Moreover, we have also included the exact $\mur$ and $\muf$ scale dependence 
at NNLO~\cite{Kidonakis:2001nj} which can be constructed using renormalization group methods. 
For the time being, we have chosen a common value $\mu$ for the scales, 
and we will address the independent variation of $\mur$ and $\muf$ in a future publication.
However, based on preliminary studies we do not expect large modifications here. 
In summary, we have accounted for all numerically dominant contributions and are confident 
that this provides a very good approximation to the unknown full NNLO result as
experience from other reactions, e.g. Higgs-production in gluon fusion~\cite{Moch:2005ky} shows.

\begin{figure}[tbp]
\begin{center}
\includegraphics[width=7.0cm]{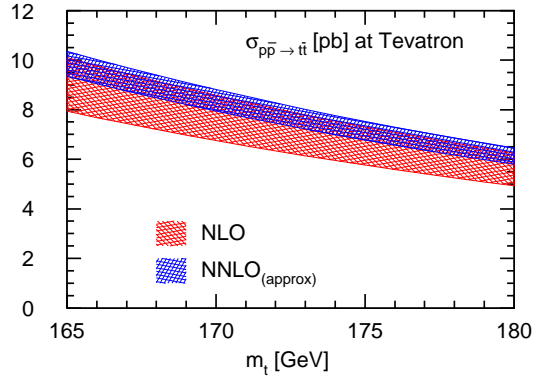}
\end{center}
\vspace*{-10mm}
\caption{ \small
  \label{fig:tev}
  The \NLO\ and \NNLO\ QCD prediction for the $t{\bar t}$ total cross section at
  Tevatron for $\sqrt{s}=1.96$~TeV.
  The bands denote the total uncertainty from PDF and scale variations for 
  the MRST06nnlo set~\cite{Martin:2007bv} according to
  Eq.~(\ref{eq:range}).
}
\end{figure}

\begin{figure}[tbp]
\begin{center}
\includegraphics[width=7.0cm]{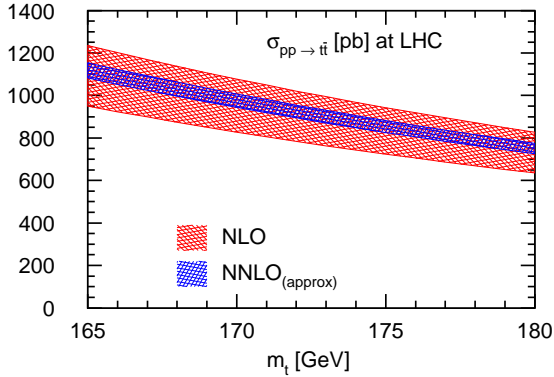}
\end{center}
\vspace*{-10mm}
\caption{ \small 
  \label{fig:lhc}
  Same as Fig.~\ref{fig:tev} for LHC with $\sqrt{s}=14$~TeV.
 }
\end{figure}

Let us next present our results for Tevatron and LHC.
In Figs.~\ref{fig:tev} and \ref{fig:lhc} we plot the uncertainty range~(\ref{eq:range}) 
comparing NLO and NNLO accuracy. 
At Tevatron (Fig.~\ref{fig:tev}) the central value at NNLO increases typically by 8\% with respect to NLO. 
The residual scale dependence of \sigmaNNLO\ is 3\%, which corresponds to a reduction by a factor of two compared to NLO.
The overall uncertainty according to Eq.~(\ref{eq:range}) is at \NNLO\ about 8\% for the
CTEQ6.6 and 6\% for the MRST06nnlo PDF set.
At LHC (Fig.~\ref{fig:lhc}) our \sigmaNNLO\ leads only to a small shift of a few
percent in the central value and the \NNLO\ band is about 6\% for CTEQ6.6
and about 4\% for MRST06nnlo, which exhibits again a drastic reduction of the
scale uncertainty as compared to the prediction based on NLO QCD.

For phenomenological applications, the results of Eqs.~(\ref{eq:totalcrs}), (\ref{eq:range}) 
are best presented by means of simple formulae for the mass dependence of the total cross section. 
To that end we make the ansatz following Ref.~\cite{Cacciari:2008zb} 
\begin{eqnarray}
  \label{eq:mass-dep}
  \sigma(\mt) = a + b x + c x^2 + d x^3 + e x^4
  \, ,
\end{eqnarray}
where $x=(\mt/\mbox{GeV}-171)$. 
The parameters $a,b,c,d,e$ are fitted 
to reproduce $\sigma$ in the mass range $150~\mbox{GeV} \le \mt \le 190~\mbox{GeV}$ 
with a typical accuracy of better than $0.1$ per mille. 
For Tevatron and LHC the respective results for various PDF sets are given in Tabs.~\ref{tab:mass-fit-tev}--\ref{tab:mass-fit-lhc}.
Note, that Eq.(\ref{eq:mass-dep}) uses a polynomial of degree four
and also determines the parameter $a$ for the central value
$\sigma(\mt=171~\mbox{GeV})$ from the fit.

\begin{table*}[htbp]
  \begin{tabular}{|l|l|l|l|l|l|l|} \hline
    \multicolumn{2}{|c|}{Tevatron $\sqrt{s}=1.96\mbox{TeV}$}
    &a[pb]&b[pb]&c[pb] $\times 10^{2}$&d[pb] $\times 10^{5}$&e[pb] $\times 10^{7}$ \\
    \hline
    CTEQ6.5
  & \central   & 7.93923 & - 0.247233  & 0.43143  & - 6.09398  & 7.37166  \\ 
  & \pdfdown   & 7.49359 & - 0.231436  & 0.39979  & - 5.59529  & 6.73356  \\ 
  & \pdfup     & 8.38488 & - 0.263031  & 0.46307  & - 6.59268  & 8.00976  \\ 
  & \highscale & 7.6572 & - 0.239371  & 0.418793  & - 5.92369  & 7.16791  \\ 
  & \lowscale  & 8.00918 & - 0.247709  & 0.428962  & - 6.00414  & 7.2013  \\ 

    \hline
    CTEQ6.6
  & \central   & 7.80984 & - 0.243547  & 0.425424  & - 6.01643  & 7.29114  \\ 
  & \pdfdown   & 7.36172 & - 0.22749  & 0.39309  & - 5.5075  & 6.6412  \\ 
  & \pdfup     & 8.25797 & - 0.259604  & 0.457757  & - 6.52536  & 7.94108  \\ 
  & \highscale & 7.53093 & - 0.235746  & 0.412856  & - 5.84596  & 7.08203  \\ 
  & \lowscale  & 7.87933 & - 0.244019  & 0.422956  & - 5.92653  & 7.11881  \\ 

    \hline
    MRST06nnlo
  & \central   & 8.24268 & - 0.262604  & 0.46727  & - 6.70602  & 8.19512  \\ 
  & \pdfdown   & 8.02331 & - 0.254847  & 0.451872  & - 6.46408  & 7.88561  \\ 
  & \pdfup     & 8.46205 & - 0.27036  & 0.482669  & - 6.94796  & 8.50462  \\ 
  & \highscale & 7.90318 & - 0.252688  & 0.450793  & - 6.47936  & 7.91184  \\ 
  & \lowscale  & 8.32601 & - 0.263279  & 0.464701  & - 6.60993  & 8.01315  \\ 

    \hline
  \end{tabular}  
  \vspace*{3mm}
  \caption{ \small
    \label{tab:mass-fit-tev}
    The coefficients of the parameterization~(\ref{eq:mass-dep}) for the 
    cross section $\sigmaNNLO$ of Ref.~\cite{Moch:2008qy} in pb at 
    Tevatron ($\sqrt{s}=1.96$~TeV) using the 
    PDF sets CTEQ6.5~\cite{Tung:2006tb}, CTEQ6.6~\cite{Nadolsky:2008zw} and 
    MRST06nnlo set~\cite{Martin:2007bv}.
  }
  \vspace*{3mm}
\end{table*}

\begin{table*}[htbp]
  \begin{tabular}{|l|l|l|l|l|l|l|} \hline
    \multicolumn{2}{|c|}{LHC $\sqrt{s}=10\mbox{TeV}$}
    &a[pb]&b[pb]&c[pb]&d[pb] $\times 10^{2}$&e[pb] $\times 10^{5}$ \\
    \hline
    CTEQ6.5
  & \central   & 418.588 & - 11.8988  & 0.19815  & - 0.267839  & 3.07947  \\ 
  & \pdfdown   & 399.508 & - 11.4849  & 0.192657  & - 0.261487  & 3.01046  \\ 
  & \pdfup     & 437.668 & - 12.3126  & 0.203644  & - 0.274191  & 3.14848  \\ 
  & \highscale & 413.69 & - 11.7539  & 0.195659  & - 0.264393  & 3.03946  \\ 
  & \lowscale  & 400.724 & - 11.3609  & 0.188716  & - 0.254368  & 2.91524  \\ 

    \hline
    CTEQ6.6
  & \central   & 419.062 & - 11.9351  & 0.199035  & - 0.269327  & 3.09784  \\ 
  & \pdfdown   & 400.4 & - 11.5275  & 0.193785  & - 0.263554  & 3.03731  \\ 
  & \pdfup     & 438.121 & - 12.3427  & 0.204284  & - 0.275101  & 3.15837  \\ 
  & \highscale & 413.944 & - 11.7839  & 0.196439  & - 0.265736  & 3.05688  \\ 
  & \lowscale  & 401.139 & - 11.3946  & 0.18954  & - 0.255779  & 2.93453  \\ 

    \hline
    MRST06nnlo
  & \central   & 449.93 & - 12.6551  & 0.208761  & - 0.279652  & 3.1917  \\ 
  & \pdfdown   & 441.788 & - 12.4928  & 0.206866  & - 0.277817  & 3.17472  \\ 
  & \pdfup     & 458.073 & - 12.8175  & 0.210656  & - 0.281487  & 3.20868  \\ 
  & \highscale & 443.176 & - 12.4618  & 0.205521  & - 0.275247  & 3.14118  \\ 
  & \lowscale  & 431.685 & - 12.111  & 0.199286  & - 0.266217  & 3.03056  \\ 

    \hline
  \end{tabular}
  \vspace*{3mm}
  \caption{ \small
    \label{tab:mass-fit-lhc10}
    Same as in Tab.~\ref{tab:mass-fit-tev} for the 
    cross section $\sigmaNNLO$ at LHC 
    with start-up energy $\sqrt{s}=10$~TeV.
  }
  \vspace*{3mm}
%
  \begin{tabular}{|l|l|l|l|l|l|l|} \hline
    \multicolumn{2}{|c|}{LHC $\sqrt{s}=14\mbox{TeV}$}
    &a[pb]&b[pb]&c[pb]&d[pb] $\times 10^{2}$&e[pb] $\times 10^{5}$ \\
    \hline
    CTEQ6.5
  & \central   & 917.522 & - 24.8588  & 0.398405  & - 0.520463  & 5.82374  \\ 
  & \pdfdown   & 887.489 & - 24.2054  & 0.388932  & - 0.507626  & 5.66159  \\ 
  & \pdfup     & 947.556 & - 25.5122  & 0.407878  & - 0.5333  & 5.98589  \\ 
  & \highscale & 908.276 & - 24.5978  & 0.394104  & - 0.514815  & 5.76339  \\ 
  & \lowscale  & 878.259 & - 23.7272  & 0.379206  & - 0.493949  & 5.51362  \\ 

    \hline
    CTEQ6.6
  & \central   & 920.475 & - 24.9757  & 0.400681  & - 0.523783  & 5.85946  \\ 
  & \pdfdown   & 891.302 & - 24.3671  & 0.392303  & - 0.512828  & 5.72046  \\ 
  & \pdfup     & 949.648 & - 25.5842  & 0.409059  & - 0.534738  & 5.99846  \\ 
  & \highscale & 910.798 & - 24.7036  & 0.396235  & - 0.517914  & 5.79166  \\ 
  & \lowscale  & 881.04 & - 23.8377  & 0.381364  & - 0.497136  & 5.54881  \\ 

    \hline
    MRST06nnlo
  & \central   & 969.128 & - 25.9797  & 0.412622  & - 0.534539  & 5.93782  \\ 
  & \pdfdown   & 957.674 & - 25.7648  & 0.409906  & - 0.530883  & 5.88547  \\ 
  & \pdfup     & 980.582 & - 26.1945  & 0.415337  & - 0.538195  & 5.99016  \\ 
  & \highscale & 956.58 & - 25.636  & 0.407062  & - 0.527262  & 5.86177  \\ 
  & \lowscale  & 929.84 & - 24.8578  & 0.393723  & - 0.508485  & 5.63364  \\ 

    \hline
  \end{tabular}
  \vspace*{3mm}
  \caption{ \small
    \label{tab:mass-fit-lhc}
    Same as in Tab.~\ref{tab:mass-fit-tev} for the 
    cross section $\sigmaNNLO$ at LHC with $\sqrt{s}=14$~TeV.
  }
  \vspace*{3mm}
\end{table*}

Finally, we briefly quote some \NNLO\ rates for the pair-production of new heavy quarks 
in the fundamental representation of the color $SU(3)$ gauge group 
at LHC with $\sqrt{s}=14$~TeV (see also Ref.~\cite{Cacciari:2008zb}).
Such particles with a mass $m_T$ appear in certain extensions of the Standard
Model and we focus on a production model which is entirely dominated by QCD effects. 
Thus, our cross section $\sigmaNNLO$ provides a meaningful and accurate prediction 
because its numerical values arises largely from the threshold region where
the logarithms $\ln^k \beta$ dominate.

In Tabs.~\ref{tab:new-hvq-mrst06}, \ref{tab:new-hvq-cteq65} we quote the corresponding numbers 
in the mass range $0.5~\mbox{TeV} \le m_T \le 2~\mbox{TeV}$ 
(see Ref.~\cite{Cacciari:2008zb} for results to NLO accuracy).
We observe that the scale dependence at NNLO accuracy is rather small, showing the
expected good stability of the perturbative prediction.
The  relative variation of $\sigma$ with respect to the PDFs, though, is dominating by far. 
Note there is the usual factor of two between the PDF uncertainty quoted by 
MRST06nnlo~\cite{Martin:2007bv} and the CTEQ6.5~\cite{Tung:2006tb} PDF sets 
due to the definition of the tolerance criteria in the respective fits.
The reason for the large observed PDF uncertainty is the gluon PDF being poorly
constrained in the relevant region of large momentum fraction $x \simeq 0.1 \dots 0.3$.
This is a fact well-known to influence many searches for high-mass
particles in gluon fusion channels  
(see e.g. Ref.~\cite{Nadolsky:2008zw} for the correlation of top-quark pair
production rate with the high mass Higgs cross section).

\begin{table*}[htbp]
\begin{tabular}{c|llc|llc|llc} \hline
 &\multicolumn{3}{|c|}{only scale uncertainty}&\multicolumn{3}{|c|}{only pdf uncertainty}&\multicolumn{3}{|c}{total uncertainty}\\ 
$m_T$[TeV] &min&max&$\delta [\%]$&min&max&$\delta [\%]$&min&max&$\delta [\%]$\\ \hline\hline
  0.5  &  4345.  &  4472.  &  1  &  4287.  &  4656.  &   4  &  4160.  &  4656.  &  6 \\
  0.6  &  1561.  &  1601.  &  1  &  1526.  &  1676.  &   5  &  1486.  &  1676.  &  6 \\
  0.7  &  634.1  &  649.2  &  1  &  616.0  &  682.5  &   5  &  600.8  &  682.5  &  6 \\
  0.8  &  282.3  &  288.5  &  1  &  272.6  &  304.4  &   6  &  266.4  &  304.4  &  7 \\
  0.9  &  134.5  &  137.2  &  1  &  129.3  &  145.1  &   6  &  126.6  &  145.1  &  7 \\
  1.0  &  67.64  &  68.94  &  1  &  64.81  &  73.08  &   6  &  63.50  &  73.08  &  7 \\
  1.1  &  35.45  &  36.17  &  1  &  33.93  &  38.41  &   6  &  33.22  &  38.41  &  7 \\
  1.2  &  19.23  &  19.65  &  1  &  18.38  &  20.91  &   6  &  17.97  &  20.91  &  8 \\
  1.3  &  10.74  &  10.99  &  1  &  10.26  &  11.72  &   7  &  10.01  &  11.72  &  8 \\
  1.4  &  6.147  &  6.301  &  1  &  5.862  &  6.741  &   7  &  5.708  &  6.741  &  8 \\
  1.5  &  3.589  &  3.687  &  1  &  3.417  &  3.957  &   7  &  3.319  &  3.957  &  9 \\
  1.6  &  2.130  &  2.192  &  1  &  2.021  &  2.363  &   8  &  1.959  &  2.363  &  9 \\
  1.7  &  1.282  &  1.322  &  2  &  1.212  &  1.432  &   8  &  1.172  &  1.432  &  10 \\
  1.8  &  0.781  &  0.806  &  2  &  0.735  &  0.878  &   9  &  0.710  &  0.878  &  11 \\
  1.9  &  0.480  &  0.497  &  2  &  0.450  &  0.544  &   9  &  0.433  &  0.544  &  11 \\
  2.0  &  0.298  &  0.309  &  2  &  0.277  &  0.340  &  10  &  0.266  &  0.340  &  12 \\
\hline
\end{tabular}
  \vspace*{3mm}
  \caption{ \small
    \label{tab:new-hvq-mrst06}
    The \NNLO\ cross section of Ref.~\cite{Moch:2008qy} in fb 
    for the pair-production of a (new) heavy quark with mass $m_T$ at LHC 
    ($\sqrt{s}=14$~TeV) using the MRST06nnlo PDF set~\cite{Martin:2007bv}.
    $\delta$ is the relative uncertainty with respect 
    to the central value: $\delta = 100\times 
    (\mbox{max}-\mbox{min})/(\mbox{max}+\mbox{min})$.
  }
  \vspace*{3mm}
%
\begin{tabular}{c|llc|llc|llc} \hline
 &\multicolumn{3}{|c|}{only scale uncertainty}&\multicolumn{3}{|c|}{only pdf uncertainty}&\multicolumn{3}{|c}{total uncertainty}\\ 
$m_T$[TeV] &min&max&$\delta [\%]$&min&max&$\delta [\%]$&min&max&$\delta [\%]$\\ \hline\hline
  0.5  &  3921.  &  4037.  &  1  &  3639.  &  4436.  &  10  &  3522.  &  4436.  &  11 \\
  0.6  &  1402.  &  1440.  &  1  &  1275.  &  1604.  &  11  &  1238.  &  1604.  &  13 \\
  0.7  &  568.0  &  582.5  &  1  &  508.6  &  656.5  &  13  &  494.1  &  656.5  &  14 \\
  0.8  &  252.4  &  258.2  &  1  &  222.5  &  294.0  &  14  &  216.6  &  294.0  &  15 \\
  0.9  &  120.3  &  122.8  &  1  &  104.5  &  141.0  &  15  &  102.0  &  141.0  &  16 \\
  1.0  &  60.48  &  61.66  &  1  &  51.94  &  71.37  &  16  &  50.76  &  71.37  &  17 \\
  1.1  &  31.78  &  32.30  &  1  &  26.95  &  37.64  &  17  &  26.44  &  37.64  &  17 \\
  1.2  &  17.25  &  17.57  &  1  &  14.50  &  20.57  &  17  &  14.21  &  20.57  &  18 \\
  1.3  &  9.626  &  9.802  &  1  &  8.023  &  11.58  &  18  &  7.847  &  11.58  &  19 \\
  1.4  &  5.503  &  5.614  &  1  &  4.555  &  6.673  &  19  &  4.444  &  6.673  &  20 \\
  1.5  &  3.208  &  3.277  &  1  &  2.630  &  3.925  &  20  &  2.560  &  3.925  &  21 \\
  1.6  &  1.902  &  1.946  &  1  &  1.545  &  2.347  &  21  &  1.502  &  2.347  &  22 \\
  1.7  &  1.144  &  1.173  &  1  &  0.921  &  1.426  &  22  &  0.892  &  1.426  &  23 \\
  1.8  &  0.696  &  0.715  &  1  &  0.554  &  0.876  &  23  &  0.535  &  0.876  &  24 \\
  1.9  &  0.428  &  0.441  &  1  &  0.337  &  0.544  &  24  &  0.324  &  0.544  &  25 \\
  2.0  &  0.265  &  0.274  &  2  &  0.206  &  0.342  &  25  &  0.198  &  0.342  &  27 \\
\hline
\end{tabular}
  \vspace*{3mm}
  \caption{ \small
    \label{tab:new-hvq-cteq65}
    Same as in Tab.~\ref{tab:new-hvq-mrst06} using the 
    CTEQ6.5~\cite{Tung:2006tb} PDF set.
  }
  \vspace*{3mm}
\end{table*}

\section{Conclusion}
\label{sec:concl}
We have presented updated predictions for cross sections of top-quark pair
production based on the (approximate) NNLO results of Ref.~\cite{Moch:2008qy}.
These represent the best present estimates for hadro-production of top-quark
pairs, both at Tevatron and LHC. 
We have argued that the neglected contributions 
(i.e. power suppressed terms away from threshold and new parton channels) are
numerically small.
We have found good convergence properties of the higher order corrections and
greatly improved stability of the total cross section with respect to scale variations by our \NNLO\ result.
For applications, we have presented simple formulae~(\ref{eq:mass-dep}) with 0.1 per mille accuracy for the mass dependence
of the total cross section in the range $150~\mbox{GeV} \le \mt \le 190~\mbox{GeV}$.
Finally, we have applied our results to estimate the pair-production rates of new quarks 
heavier than the top-quark in the range up to $2~\mbox{TeV}$.

The results of Tabs.~\ref{tab:mass-fit-tev}--\ref{tab:mass-fit-lhc} 
for the fit of the mass dependence of $\sigma$ have
also been coded in a C-program, which is available from the authors upon request.

\subsection*{Acknowledgments}
S.M. is supported by the Helmholtz Gemeinschaft under contract VH-NG-105 and 
and P.U. is a Heisenberg fellow of Deutsche Forschungsgemeinschaft (DFG).
This work is also partly supported by DFG in SFB/TR 9.


\begin{thebibliography}{10}

\bibitem{Bernreuther:2008ju}
W. Bernreuther,
\newblock J. Phys. G35 (2008) 083001, arXiv:0805.1333 [hep-ph]
\newblock 

\bibitem{Nason:1988xz}
P. Nason, S. Dawson and R.K. Ellis,
\newblock Nucl. Phys. B303 (1988) 607
\newblock 

\bibitem{Beenakker:1989bq}
W. Beenakker et~al.,
\newblock Phys. Rev. D40 (1989) 54
\newblock 

\bibitem{Bernreuther:2004jv}
W. Bernreuther et~al.,
\newblock Nucl. Phys. B690 (2004) 81, hep-ph/0403035
\newblock 

\bibitem{Dittmaier:2007wz}
S. Dittmaier, P. Uwer and S. Weinzierl,
\newblock Phys. Rev. Lett. 98 (2007) 262002, hep-ph/0703120
\newblock 

\bibitem{Czakon:2007ej}
M. Czakon, A. Mitov and S. Moch,
\newblock Phys. Lett. B651 (2007) 147, arXiv:0705.1975 [hep-ph]
\newblock 

\bibitem{Czakon:2007wk}
M. Czakon, A. Mitov and S. Moch,
\newblock Nucl. Phys. B798 (2008) 210, arXiv:0707.4139 [hep-ph]
\newblock 

\bibitem{Korner:2008bn}
J.G. K\"orner, Z. Merebashvili and M. Rogal,
\newblock (2008), arXiv:0802.0106 [hep-ph]
\newblock 

\bibitem{Czakon:2008zk}
M. Czakon,
\newblock Phys. Lett. B664 (2008) 307, arXiv:0803.1400 [hep-ph]
\newblock 

\bibitem{Moch:2008qy}
S. Moch and P. Uwer,
\newblock Phys. Rev. D (2008) in press, arXiv:0804.1476 [hep-ph]
\newblock 

\bibitem{Hagiwara:2008df}
K. Hagiwara, Y. Sumino and H. Yokoya,
\newblock (2008), arXiv:0804.1014 [hep-ph]
\newblock 

\bibitem{Kidonakis:2001nj}
N. Kidonakis et~al.,
\newblock Phys. Rev. D64 (2001) 114001, hep-ph/0105041
\newblock 

\bibitem{Kidonakis:2008mu}
N. Kidonakis and R. Vogt,
\newblock (2008), arXiv:0805.3844 [hep-ph]
\newblock 

\bibitem{Cacciari:2008zb}
M. Cacciari et~al.,
\newblock (2008), arXiv:0804.2800 [hep-ph]
\newblock 

\bibitem{Nadolsky:2008zw}
P.M. Nadolsky et~al.,
\newblock (2008), arXiv:0802.0007 [hep-ph]
\newblock 

\bibitem{Martin:2007bv}
A.D. Martin et~al.,
\newblock Phys. Lett. B652 (2007) 292, arXiv:0706.0459 [hep-ph]
\newblock 

\bibitem{Tung:2006tb}
W.K. Tung et~al.,
\newblock JHEP 02 (2007) 053, hep-ph/0611254
\newblock 

\bibitem{Moch:2005ky}
S. Moch and A. Vogt,
\newblock Phys. Lett. B631 (2005) 48, hep-ph/0508265
\newblock 

\end{thebibliography}
\end{document}